\documentstyle[11pt,newpasp,twoside,epsf]{article}
\def\edcomment#1{\iffalse\marginpar{\raggedright\sl#1\/}\else\relax\fi}
\reversemarginpar

\begin{document}
\title{Further Analysis of Stellar Magnetic Cycle Periods}
\author{Steven Saar}
\affil{Harvard-Smithsonian Center for Astrophysics, 60 Garden Street,
Cambridge, MA 02138, USA}
\author{Axel Brandenburg}
\affil{NORDITA, Blegdamsvej 17, DK-2100 Copenhagen \O, Denmark} 

\def\Ro{{\rm R}_o}
\def\Rm{{\rm R}_m}
\def\oc{$\omega_{\rm cyc}$}
\def\pc2{$P_{\rm rot}^{(2)}$}
\def\Rstar{R_\ast}
\def\tauc{\tau_c}
\def\uv#1{{\hat{\bf e}}_{#1}}

\begin{abstract}
We further investigate relationships between activity cycle periods 
in cool stars and rotation to include new 
cycle data, and explore different parameterizations of the problem. 
We find that relations between cycle and rotational frequencies 
(\oc~vs. $\Omega$) and between their ratio and the inverse Rossby number
(\oc/$\Omega$ vs. Ro$^{-1}$) show many similarities, including three branches
and similar rms scatter.  We briefly discuss some implications for dynamo 
models.
\end{abstract}

\section{Introduction}\label{Intro}

Several recent studies (Ossendrijver 1998; Tobias 1998; Brandenburg et al. 1998;
Saar \& Brandenburg 1999 [=SB]; Lanza \& Rodon\`o 1999) have 
revisited relationships
between stellar magnetic cycles and other stellar properties, taking
advantage of the increased quality and quality of the cycle data available
(e.g., Baliunas et al. 1995). SB studied relationships between non-dimensional
quantities such as cycle-to-rotational frequency ratio \oc/$\Omega$,
the normalized Ca {\sc II} HK emission flux $R'_{\rm HK}$, and the inverse
Rossby number Ro$^{-1} = 2\tau_c\Omega$ (where
$\tau_c$ is the convective turnover time).  They found evidence for three
power-law ``branches" upon which stars tended to cluster.  Here we expand on
this work. We add new cycle data, and investigate how the new data affect
various parameterizations, both dimensional and non-dimensional, of
the stellar cycles, focusing on relations between \oc~and rotation.

\section{Data and Analysis}

We combine cycle and stellar data gathered by SB with more recent measurements
of plage (e.g., Hatzes et al. 2000) and spot cycles (e.g., Ol\'ah et al. 2000).
Cyclic changes in $P_{\rm rot}$ in some close binaries
have been linked with magnetic cycle modulation (via changes in
mean magnetic pressure) of stellar quadrupole moments (Lanza et al. 1998).
These cycles based on variations in $P_{\rm rot}$ (Lanza \& Rodon\`o 1999)
are also tentatively included.  We follow the strategy of SB, using theoretical
$\tau_c$ (Gunn et al. 1998) and weighting the $P_{\rm cyc}$ by a ``quality
factor" $w$ ($0.5 \leq w \leq 4$)
depending on the strength of the periodogram signal or
clarity of the cycle modulation.  We set $w=1$ for
the $P_{\rm rot}$-change cycles.  Stars are assigned to branches (where
appropriate) by eye to minimize fit rms.  Evolved stars were not included
in the fits due to less well determined $\tau_c$.  Results for 
different
classes of stars are shown in Fig. 1 LEFT (using dimensionless
\oc/$\Omega$ and Ro$^{-1}$) and Fig. 1  RIGHT (using \oc~=$2\pi/P_{\rm cyc}$ 
and $\Omega = 2\pi/P_{\rm rot}$).

\begin{figure*}[th]

\plottwo{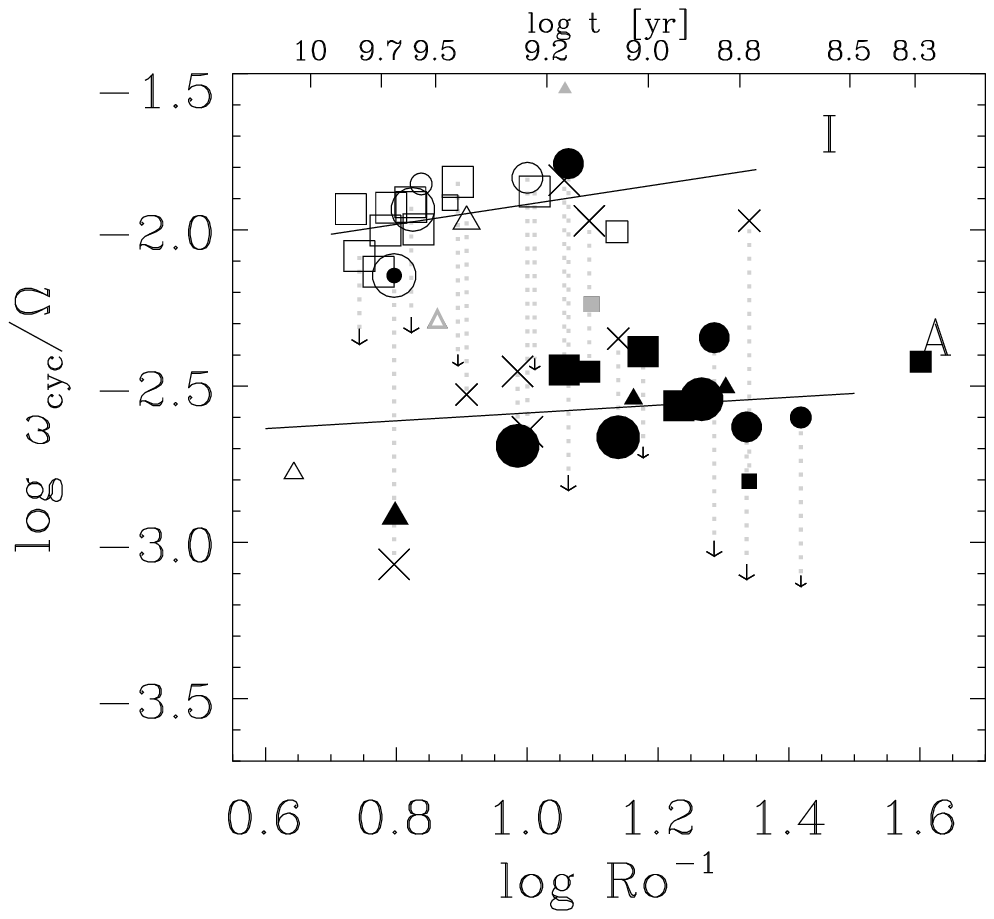}{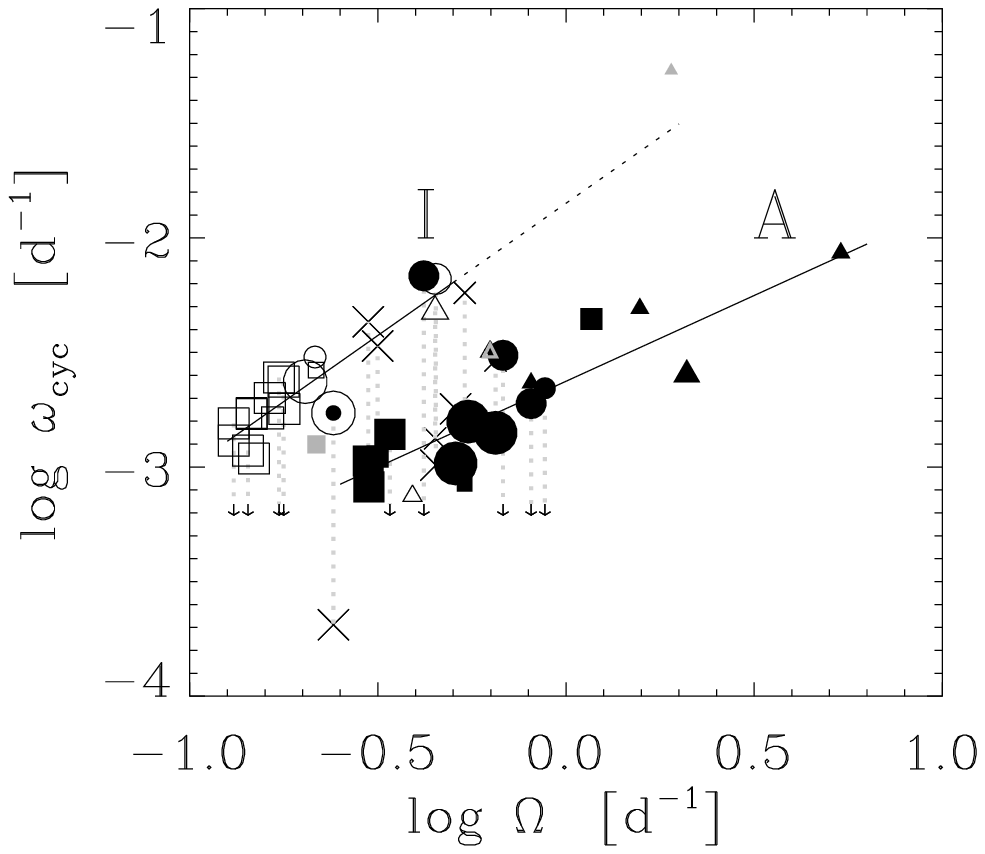}

\plottwo{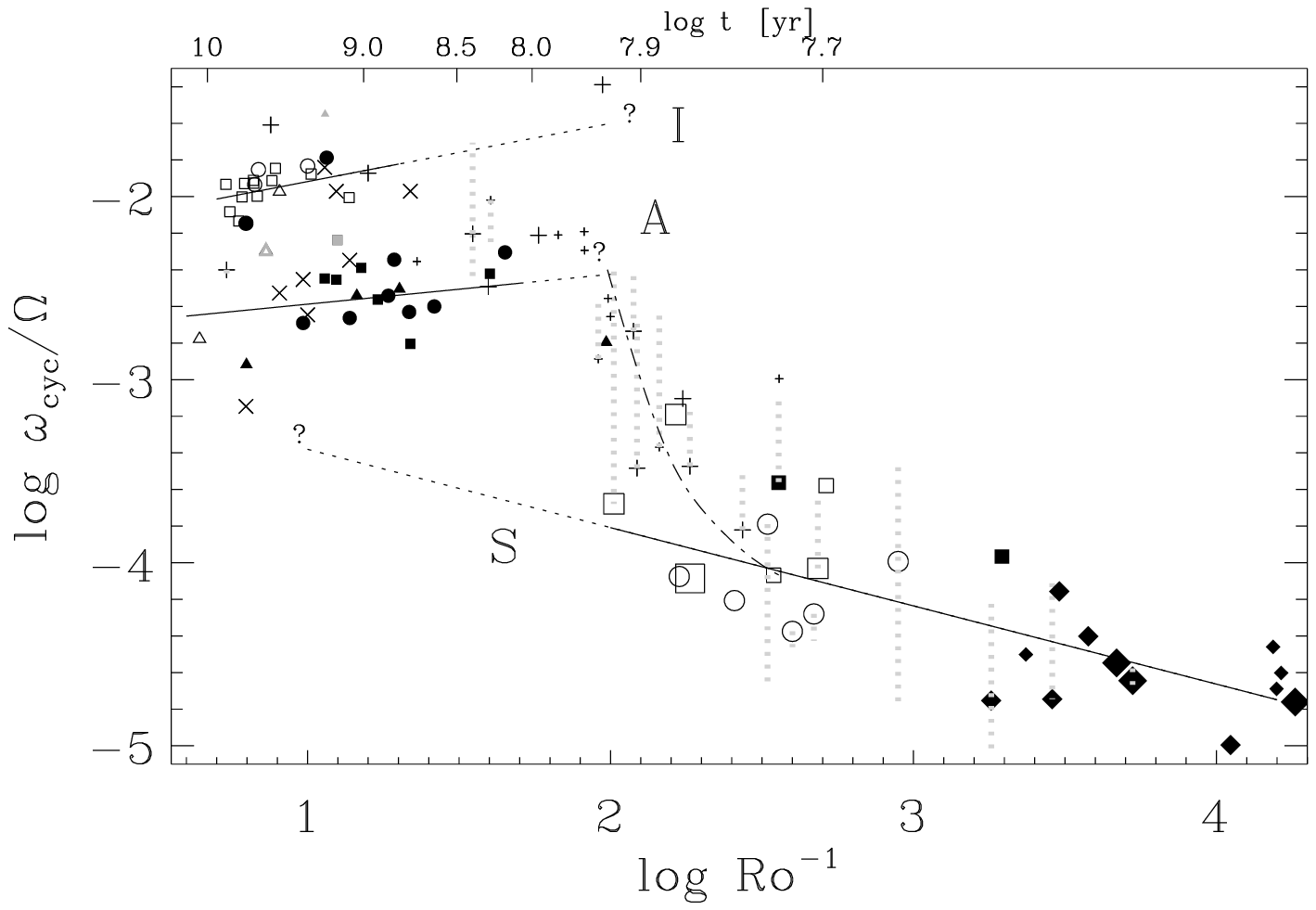}{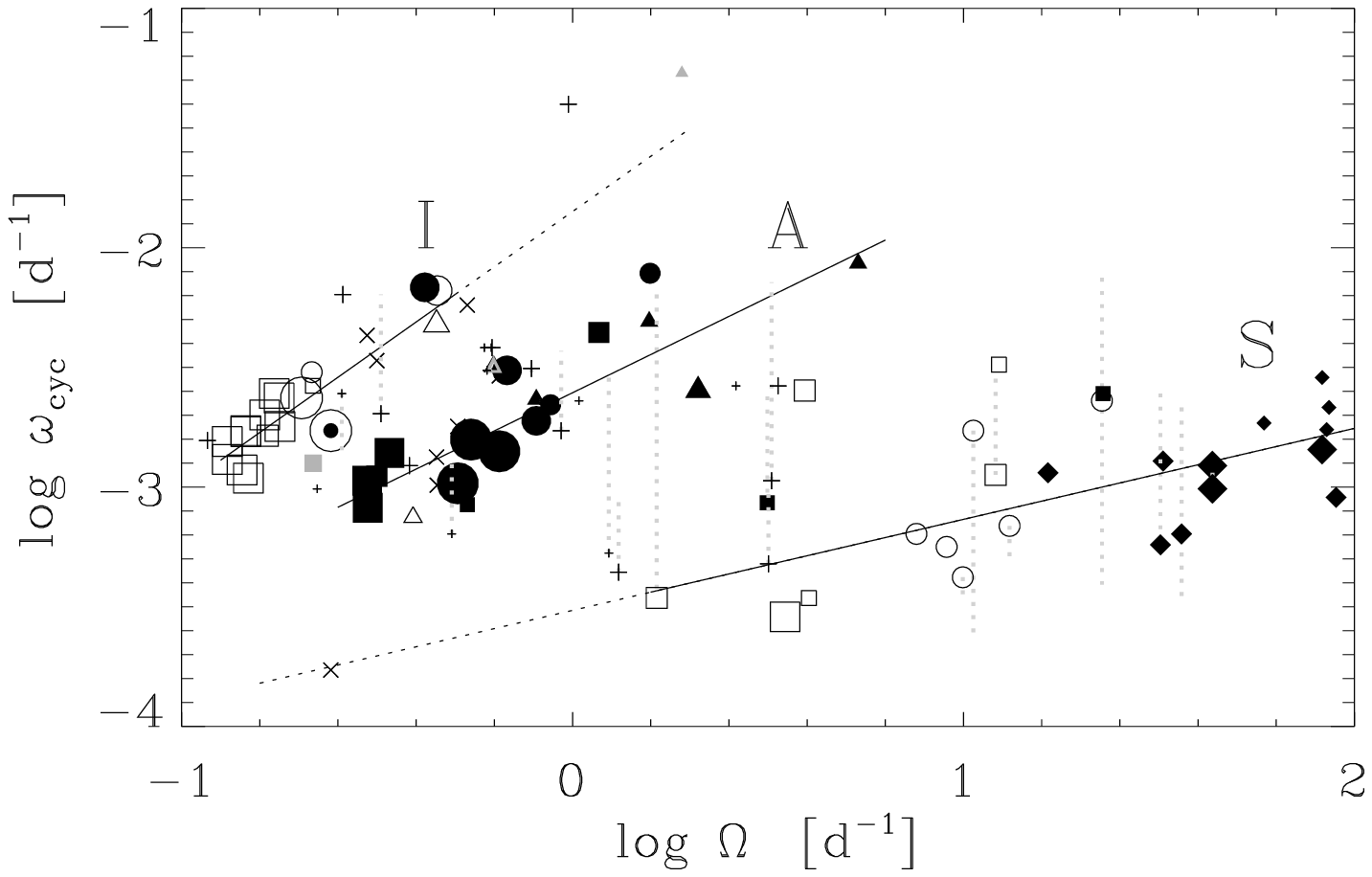}

\plottwo{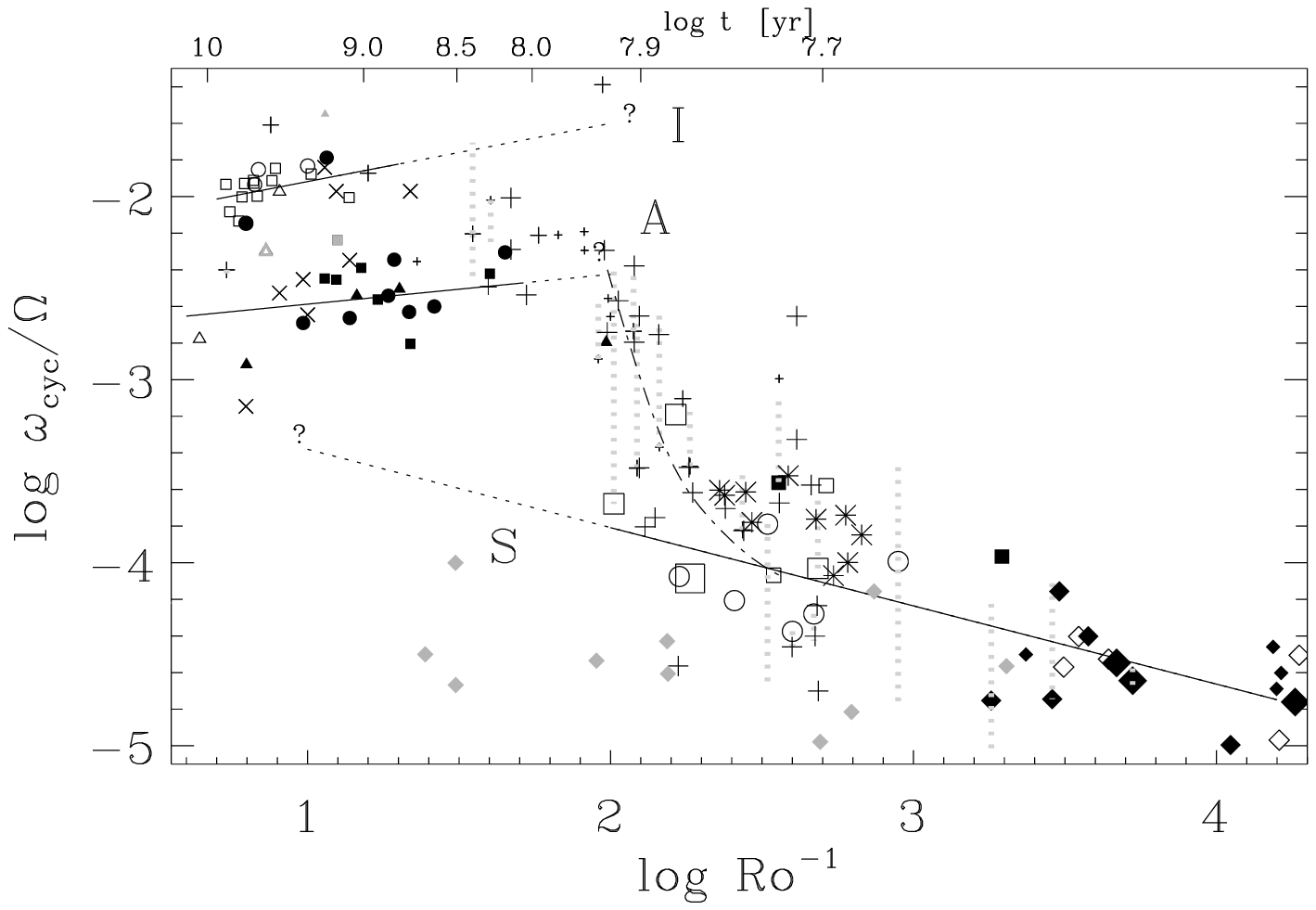}{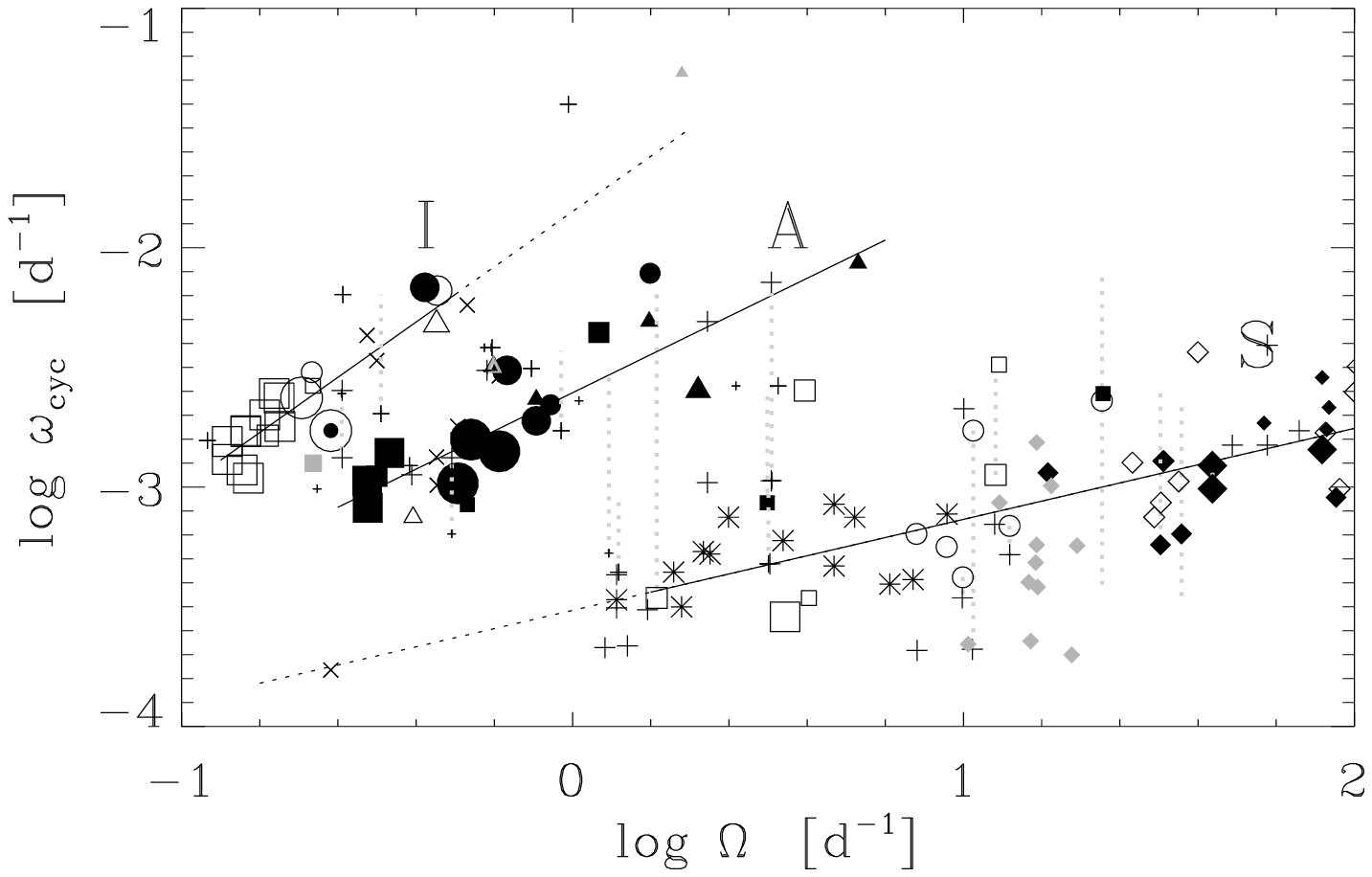}

\caption{
\small
LEFT \underline{top:}
\oc/$\Omega$ vs. Ro$^{-1}$ for single dwarfs; symbols indicate
the sun ($\odot$), F ($\triangle$), G ($\circ$), and K ($\Box$) stars
(filled if log $R'_{\rm HK} \geq -4.75$; 
size $\propto \sqrt{w}$, the $P_{\rm cyc}$ ``reliability"). 
Dotted vertical lines connect two $P_{\rm cyc}$
(a $\times$ marks \pc2), or $P_{\rm cyc}$
with a long-term trend (i.e., a possible \pc2 $> 25$ yr;
arrow symbol).  Weighted least square fits (\oc/$\Omega \propto$
Ro$^{\delta}$) for the active (A) and inactive (I) branches are shown
(solid); $\delta_I = -0.32$ and $\delta_A = -0.16$. 
LEFT \underline{middle:}
same, including binaries (BY Dra, CV secondaries;
M stars = $\diamond$) and RS CVns ($+$; not included in the fits).
A ``superactive" (S) branch appears, with $\delta_S = +0.43$ (\pc2 
$\leftrightarrow P_{\rm cyc}$  
lines shown only for new stars).
A transitional regime between the A and S branches
is indicated (dash-dot).
LEFT \underline{bottom:}
same, including cycles based on $P_{\rm rot}$ variation in
RS CVns (new $+$), CV secondaries (open $\diamond$), Algols ($*$), and
contact binaries (gray $\diamond$).  
RIGHT \underline{top:}
\oc~vs. $\Omega$ for single dwarfs.
Fits (\oc $\propto \Omega^\delta$) for the
A and I branches (solid) yield 
$\delta_I = 1.15$ and $\delta_A = 0.80$.
RIGHT \underline{middle:}
same, including binaries (like LEFT middle). 
The new S branch shows $\delta_S = 0.38$. 
RIGHT \underline{bottom:}
same, including $P_{\rm rot}$ variation cycles (like LEFT bottom).
\normalsize
}
\label{Fig1}
\end{figure*}

\section{Results and Discussion}

Our results can be summarized as follows:

{\bf (1)} Three branches -- denoted I (inactive), A (active), and S
(super-active) -- appear in both the Ro$^{-1}$ and $\Omega$
parameterizations (Fig. 1).  For the Ro$^{-1}$ fit, the
power law exponents are $\delta_I \approx -0.3$ (with a 
fit dispersion $\sigma_{fit} = 0.095$ dex), 
$\delta_A \approx -0.15$ ($\sigma_{fit} =  0.18$),
and $\delta_S \approx 0.4$ ($\sigma_{fit} = 0.26$ dex); for the $\Omega$ fit,
$\delta_I \approx 1.15$ ($\sigma_{fit} = 0.093$ dex), $\delta_A \approx 0.8$
($\sigma_{fit} = 0.17$ dex),
and $\delta_S \approx 0.4$ ($\sigma_{fit} = 0.24$ dex).  
Thus the rms scatter is similar for the two parameterizations.

{\bf (2)} Secondary cycle periods (\pc2) 
seen in some stars often lie on one of the
branches (though this is more 
rare in S branch stars).  The solar Gleissberg ``cycle"
($\sim 100$ years) appears to lie on the S branch.  
The preferred branch of the primary
$P_{\rm cyc}$  (with the strongest periodogram signal) may be mass
and $\Omega$ dependent.  Multiple $P_{\rm cyc}$ may
reflect multiple dynamo modes in an $\alpha\Omega$ framework
(Knobloch, Rosner \& Weiss 1981), or 
different dynamos existing in separate latitude zones
(note the dual, separately evolving activity patterns in the 
double $P_{\rm cyc}$ 
star $\beta$ Comae; Donahue \& Baliunas 1992).
In the Babcock-Leighton scenario, \pc2~may 
be excited by stochastic variations in the poloidal source term 
(Charbonneau \& Dikpati 2000).

{\bf (3)} A single power law can be fit to the data
(e.g., \oc $\propto \Omega^{-0.09}$, SB;  see also Baliunas et al. 1996) 
but only at the expense of a 
considerably higher dispersion about the fit ($\sigma_{fit} = 0.33$ dex),
and loss of an explanation for the secondary cycle periods
(since they no longer reside on another dynamo ``branch"). 

{\bf (4)} Evolved stars typically lie near branches, though show more scatter
than the dwarfs.  Since the increased scatter is seen in both 
parameterizations,
it is unlikely to be due to less precise $\tau_c$ in evolved stars
(indeed, arguably the scatter in evolved stars is reduced
using Ro$^{-1}$). The $P_{\rm cyc}$ based on $P_{\rm rot}$ 
variation (Lanza \& Rodon\`o 1999) also follow the general trends. 
The branches are better separated using Ro$^{-1}$. On the other hand, 
the $\Omega$ plot is simpler, lacking the ``transitional" regime between the
A and S branches seen in the Ro$^{-1}$ diagrams.  
Contact binaries (gray $\diamond$; bottom panels) are poorly
fit in both schemes (worse if Ro$^{-1}$ is used); their dynamos may be
altered by turbulent energy transfer toward the secondary (Hazlehurst 1985)
which is independent of rotation.

{\bf (5)} The branches may merge for small Ro$^{-1}$ or $\Omega$
(though at values which might not be reached by actual stars).
Curiously, the ratio of the power law exponents for the $\Omega$ fits are
$\delta_I:\delta_A:\delta_S \approx 3:2:1$.

{\bf (6)} Since $\Omega$ and Ro$^{-1}$ decrease in time on the main-sequence, 
the relations between
\oc~and rotation map out dynamo evolution with time.  The overlapping branches
and \pc2~suggest that \oc~evolves in time in a
complex, sometimes multi-valued fashion.  The panels of Figure 1 LEFT show an
approximate age calibration along the top axes.

A Babcock-Leighton type model predicts \oc $\propto u_m^{0.9}$
for solar-like dwarfs (where $u_m$ is the meridional flow velocity;
Dikpati \& Charbonneau 1999).  If $u_m$ increases approximately linearly 
with $\Omega$ in slower rotators (e.g., Brummell et al. 1998), 
the predicted \oc~matches the I and A branches reasonably well
(see also Charbonneau \& Saar, this volume).  Mean-field models
with sufficiently strong $\Omega$ dependence for
the differential rotation (e.g., Donahue et al. 1996) and 
the $\alpha$ effect (e.g., Brandenburg \& Schmitt 1998) 
can also match the observed branches
(SB; Charbonneau \& Saar, this volume).
We are studying a variety of dynamo models to better understand the
implications of the cycle - rotation relations seen here.

\acknowledgements
This work was supported by NSF grants AST-9528563 and AST-9731652, and
HST grants GO-7440 and GO-8143.  We thank M. Dikpati and P. Charbonneau for
enlightening discussions.


\begin{thebibliography}{}

\bibitem[]{} Baliunas, S.L., Donahue, R.A., Soon, W. et al. 1995, ApJ 438, 269 

\bibitem[]{} Baliunas, S.L., Nesme-Ribes, E., Sokoloff, D., \& Soon, W. 1996, 
ApJ 460, 848

\bibitem[]{} Brandenburg, A., Saar, S.H., \& Turpin, C.J. 1998, ApJ 498, L51

\bibitem[]{} Brandenburg, A., \& Schmitt, D. 1998, A\&A 338, L55

\bibitem[]{} Brummell, N.H., Hurlburt, N.E., \& Toomre, J. 1998, ApJ 493, 955

\bibitem[]{} Charbonneau, P. \& Dikpati, M. 2000, ApJ 543, 1027

\bibitem[]{} Dikpati, M. \&  Charbonneau, P. 1999, ApJ 518, 508

\bibitem[]{} Donahue, R.A. \& Baliunas, S.L. 1992, ApJ 393, L63 

\bibitem[]{} Donahue, R.A., Saar, S.H., \& Baliunas, S.L. 1996, ApJ 466, 384

\bibitem[]{} Gunn, A.G., Mitrou, C.K.,  \& Doyle, J.G. 1998, MNRAS 296, 150

\bibitem[]{} Hatzes, A.P., Cochran, W.D., McArthur, B. et al. 2000, ApJ 544, 145

\bibitem[]{} Hazlehurst, J. 1985, A\&A 145, 25

\bibitem[]{} Knobloch, E., Rosner, R., \& Weiss, N.O. 1981, MNRAS 197, 45

\bibitem[]{} Lanza, A., \& Rodon\`o, M. 1999, A\&A 349, 887

\bibitem[]{} Lanza, A.,  et al. 
1998, MNRAS 296, 893

\bibitem[]{} Ol\'ah, K., Koll\'ath, Z., \& Strassmeier, K.G. 2000, A\&A 356, 643

\bibitem[]{} Ossendrijver, A.J.H. 1997, A\&A 323, 151

\bibitem[]{} Saar, S.H., \& Brandenburg, A. 1999, ApJ 524, 295 [=SB]

\bibitem[]{} Tobias, S.M. 1998, MNRAS 296, 653



\end{thebibliography}
\end{document}